\documentstyle[12pt,epsf]{article}
\begin{document}
\begin{titlepage}
{}~\vspace{1.5cm}

{}~\vskip1cm
\begin{center}
\Large\bf Dispersion Relations and Rescattering Effects\\
 in B Nonleptonic Decays
\end{center}
\vspace{1.0cm}
\begin{center}
I. Caprini and L. Micu\\[0.1cm]
National Institute of Physics and Nuclear Engineering,\\
 POB MG 6, Bucharest, R-76900 Romania \\[0.3cm]
C. Bourrely\\[0.1cm]
Centre de Physique Th\'eorique 
\footnote{Laboratoire propre au CNRS-UPR 7061},CNRS-Luminy\\
Case 907, F-13288 Marseille Cedex 9 - France
\end{center}
\vspace{0.5cm}
\begin{abstract}
Recently, the final state strong interactions in nonleptonic B
decays were investigated in a formalism based on hadronic unitarity 
and dispersion relations in terms of the off-shell mass squared of 
the $B$ meson. We consider an heuristic derivation of the dispersion 
relations in the mass variables using the reduction LSZ formalism and
find a discrepancy between the  spectral function and the dispersive 
variable  used in the recent works. The part of the unitarity sum which 
describes final state interactions is shown to appear as spectral function
in a dispersion relation based on the analytic continuation in the
mass squared of one final particles. As an application, by combining this
formalism with Regge theory and $SU(3)$ flavour symmetry we obtain 
constraints on the tree and the penguin amplitudes of the decay
$B^0\to \pi^+\pi^-$.
\end{abstract}
\vspace{1.0cm}
\noindent PACS No.: 14.4, 13.25, 11.55, 11.55Fv

\noindent Number of figures: 2

\noindent March 1999

\noindent CPT-99/P.3795

\noindent Web address: www.cpt.univ-mrs.fr

\end{titlepage}
\newpage

\section{Introduction}
The final state strong 
interactions are known to play an important role in weak
nonleptonic decays. In particular,  the interplay between the strong and
the weak phases  is required by many signals of direct
$CP$ violation in $B$ decays. The presence of final state
rescattering is relevant also for the extraction of the $CKM$
phases from time dependent asymmetries, and can  modify the magnitude 
of some processes suppressed in  the Standard Model such
as $B\to \pi K$ transitions, thus reducing their sensitivity to New Physics.

The effect of the final state interactions (FSI) was
assumed until recently to be small in the very energetic
decays like those of $B$ meson to light pseudoscalar mesons, where the final
particles get apart very quickly and have no time to interact 
strongly by soft multigluon exchanges.  A related argument was based on the 
fact that at the high energy scale imposed by the mass of the $B$ meson
there is a suppression of rescattering due to the specific form of the 
Regge amplitudes dominant at this scale.  Recently, this
qualitative picture was challenged  by a more detailed dynamical approach
\cite{Dono1}-\cite{Falk}. The crucial remark made in \cite{Dono1} is that,
contrary to conventional expectations, the soft final state interactions do
not disappear at the center of mass energy set by $m_B$.
The analysis made in \cite{Dono1}-\cite{Falk} is based on hadronic
unitarity and very general features of high energy soft interactions.
Consider the weak decay 
$B \to P_1 P_2$, where $P_i$ are pseudoscalar mesons, 
and denote by $A_{B\to P_1 P_2}$ the amplitude of this process.
In the most general way, the unitarity of the $S$-matrix
allows one to express the discontinuity of  $A_{B\to P_1 P_2}$ as
\begin{equation}\label{unit}
{\mathcal D}isc A_{B\to P_1 P_2}={1\over 2i}\left[
\langle P_1 P_2| {\mathcal T}|B\rangle -
\langle P_1 P_2| {\mathcal T}^\dagger|B\rangle \right]=
{1\over 2}\sum_{I}\langle P_1 P_2| {\mathcal T}^\dagger|
I\rangle\langle I|{\mathcal T}|B\rangle \,,
\end{equation}
where  ${\mathcal T}$ is the transition operator
($S=1-i {\mathcal T}$),  which describes both the weak  and strong
interactions. To first order, the weak hamiltonian
${\mathcal H}_w$ can  appear either in the first matrix element
of the product in the right hand side of (\ref{unit}), or in the second. The intermediate
hadronic states $\{I\}$ depend of course on the place of ${\mathcal H}_w$. 
If ${\mathcal H}_w$ is acting in the matrix element containing $B$,
$\{I\}$ denote  hadronic states 
produced by the weak decay of $B$ and connected to 
the final state  $\{P_1 P_2\}$ by  a strong rescattering.  
It is this configuration of the sum (\ref{unit}) 
which describes the final state interactions in $B$ nonleptonic decays. 
Alternatively, the operator  ${\mathcal H}_w$ can be 
 located in the first matrix element in the 
unitarity sum, in which case
 $\{I\}$  are states produced by the strong decay of 
$B$, and connected to $\{P_1 P_2\}$ through a weak interaction. 
When all the particles are on-shell these terms vanish, since $B$ 
is stable with respect to strong interactions. 
The above remarks, though rather trivial, will be
useful below for the discussion of the dispersion relations.

According to general principles, the decay amplitude $A_{B\to P_1 P_2}$ 
can be obtained from its discontinuity  by means of a dispersion 
relation. This approach was considered in  \cite{Blok}-\cite{Falk}, 
where, neglecting possible subtractions, a dispersion relation  of the
following form was used
\begin{equation}\label{drel}
A(m_B^2, m_1^2, m_2^2)
={1\over \pi }\int _{s_0}^\infty {\mathrm d}s{{\mathcal D}isc A(s+i\epsilon,
m_1^2, m_2^2)
\over s-m_B^2-i\epsilon}\,.
\end{equation}
In this relation and the similar ones written below, the limit
$\epsilon\to0$ is implicitly assumed.

We use here the notation $A_{B\to P_1 P_2}= A(m_B^2, m_1^2, m_2^2)$
to show explicitely the dependence of the decay amplitude on the
external masses. As for the discontinuity (already divided by $2i$),
it was taken in   \cite{Blok}-\cite{Falk} as
the "rescattering part" of the unitarity sum (\ref{unit}) discussed above,
evaluated for an off-shell $B$ meson of mass squared equal to $s$.

The dispersion relation (\ref{drel}) is based obviously on the analytic
continuation of the decay matrix element with respect to the
mass of the initial meson $B$. We recall that dispersion relations
 in the  external mass variables were derived
in the frame of axiomatic field theory \cite{Oehme}-\cite{LSZ},
and  were used in phenomenological calculations
of form factors \cite{Binc},\cite{Omnes} (see also \cite{Bart}).
Heuristic derivations of such dispersion relations are based on the
Lehmann, Symanzik, Zimmermann (LSZ) reduction
formalism  \cite{LSZ}, combined with causality and  hadronic
unitarity. As emphasized in \cite{Bart} one must be cautious
 in using such dispersion relations, since the heuristic conjectures 
might be violated, in particular
through the appearance of anomalous thresholds.

Of interest to the present work is the fact that the specific form of 
the dispersion relation and of its discontinuity depends on which 
of the external particles is reduced in the LSZ formula.
By treating the matrix element of the decay  $B\to P_1P_2$ in the frame
of this formalism, one can prove that the spectral function of a 
dispersion relation with respect to  $s$ (the mass squared of the $B$ meson),
like Eq.~(\ref{drel}), is given {\it not} by the ``rescattering''
part of (\ref{unit}), as it was assumed in \cite{Blok}-\cite{Falk}, but by 
the second class terms in this sum.
Moreover, one can also prove that the terms describing the final
state interaction in the unitarity sum (\ref{unit})
appear as  spectral function in  a dispersion relation in terms of the mass
squared of one of the final mesons ($P_1$ or $P_2$). Therefore, the
calculation of the whole decay amplitude from its discontinuity
proceeds along a different line than that applied in
Refs.\cite{Blok}-\cite{Falk}.

In the present paper, we provide arguments for the assertions made above
and consider some applications.
In the next Section, using the LSZ formalism, we discuss the heuristic
derivation of dispersion relations for
the  decay amplitude $A_{B\to P_1 P_2}$, when either
$B$ or one of the final mesons   $P_1$ or  $P_2$ are off-mass shell. We do not
attempt to give rigorous proofs, but only to establish the correspondence
between the dispersion variable and the expression of the absorbtive part.
In Section 3 we consider in more details the approximation of  two
particle unitarity  combined with Regge theory for strong interactions,
and in Section 4 we discuss some applications: first we briefly
indicate how the conclusions of Ref. \cite{Falk} are modified by
the use of the adequate dispersion relation. Then, by combining the
formalism with $SU(3)$ flavour symmetry we derive
constraints on the amplitudes of  the $B^0\to \pi^+\pi^-$ decay. 

\section{ Dispersion relations in the external mass variable}

We consider the weak decay amplitude $A_{B\to P_1 P_2}$ defined as
\begin{equation}\label{def}
 A (p^2, k_1^2, k_2^2)= \langle P_1(k_1) P_2(k_2), out|
{\mathcal H}_w(0)|B(p)\rangle\,,
\end{equation}
where we indicated  the dependence on  the Lorentz invariants
$p^2, k_1^2$ and  $k_2^2$ (for the physical amplitude
$p^2=m_B^2,\, k_1^2=m_1^2,\, k_2^2=m_2^2$ ).
By  applying the well known LSZ formalism \cite{LSZ} we "reduce"  the particle
$P_1$, which gives
\begin{equation}\label{lsz1}
A (m_B^2, k_1^2, m_2^2)= {i\over \sqrt{2 \omega_1}}\int {\mathrm d}x
{\mathrm e}^{ik_1 x} \theta (x_0) \langle P_2(k_2)|[\eta _1(x),
{\mathcal H}_w(0)]|B(p)\rangle\,.
\end{equation}
In this relation, $\eta_1(x)$ denotes the source of the meson $P_1$, 
defined as $ {\mathcal K}_x \phi_1(x)= \eta_1(x)$,
where ${\mathcal K}_x$ is the Klein-Gordon operator and  $\phi_1(x)$ the
interpolating field ($\omega_1=\sqrt{{\mathbf k}_1^2+m_1^2}$,  
is the energy of the on shell particle $P_1$). We use
here and in what follows the fact that the single particle states
$in$ and $out$ are identical. As shown in \cite{Oehme}-\cite{LSZ}, due to
the factor $\theta(x_0)$ the
amplitude (\ref{lsz1}) can be extended as an analytic function in the
upper half of the complex plane  of the time component $k_{10}$.
A more detailed analysis \cite{KaWi}, exploiting also the causality
properties of the retarded commutator and the relation $k_1^2=k_{10}^2
+{\mathbf k}^2_1$, shows that the amplitude
$A (m_B^2, k_1^2, m_2^2)$ can be extended as an analytic function in
the whole complex plane $k_1^2$, cut along a part of the real axis, where its
discontinuity  is \cite{Oehme}
\begin{equation}\label{spectr}
{\mathcal D}isc  A(m_B^2, k_1^2, m_2^2)= {1\over 2\sqrt{2 \omega_1}}\int
{\mathrm d}x  {\mathrm e}^{ik_1 x} \langle P_2(k_2)|[\eta _1(x),
{\mathcal H}_w(0)]|B(p)\rangle\,.
\end{equation}
This discontinuity coincides actually with the imaginary part 
$\hbox {Im} A(m_B^2, k_1^2+i\epsilon, m_2^2)$ of the decay
amplitude on the upper edge of the cut (it will be shown below that this spectral 
function is real). The r.h.s. of (\ref{spectr}) is treated in the standard 
way by inserting a complete set of states in the two terms of the commutator. 
By performing the translation
\begin{equation}\label{trans}
\eta_1(x)={\mathrm e}^{iPx} \eta_1(0){\mathrm e}^{-iPx}\,,
\end{equation}
we write (\ref{spectr}) as
\begin{eqnarray}\label{spectr1}
&&\hbox{Im}  A(m_B^2, k_1^2+i\epsilon, m_2^2)= {1\over 2 \sqrt{2 \omega_1}} 
\int {\mathrm d}x  {\mathrm e}^{ik_1 x}\times \nonumber \\
&&\sum_{n}\left[{\mathrm e}^{i(k_2-p_n)x}\langle P_2(k_2)|
\eta _1|n\rangle\langle n| {\mathcal H}_w|B(p)\rangle-
{\mathrm e}^{i(p_n-p)x}\langle P_2(k_2)|
{\mathcal H}_w |n\rangle\langle n|\eta _1|B(p)\rangle\right]\,,
\end{eqnarray}
where we denoted  ${\mathcal H}_w={\mathcal H}_w(0)$ and  $\eta _1=\eta _1(0)$.
The trivial integral with respect to $x$ gives
\begin{eqnarray}\label{spectr2}
\hbox{Im}  A(m_B^2, k_1^2+i\epsilon, m_2^2)&=& {1\over 2 \sqrt{2 \omega_1}}
\sum_{n}\left[\delta  (k_1+k_2-p_n) \langle P_2(k_2)|
\eta _1|n\rangle\langle n| {\mathcal H}_w|B(p)\rangle \right. \nonumber \\
&&-\left.\delta (k_1+p_n-p)\langle P_2(k_2)|
{\mathcal H}_w |n\rangle\langle n|\eta _1|B(p)\rangle\right]\,.
\end{eqnarray}
The states contributing to the first sum have the 4-momentum $p_n=k_1
+k_2=p$ and the invariant mass $p_n^2=m_B^2$, they correspond to
what we called above the "rescattering" part of (\ref{unit}). It is
easy to see that this sum is nonzero for $k_1^2$ in
the allowed interval $0< k_1^2 < (m_B-m_2)^2$ (as we mentioned
anomalous thresholds might be present).
As the second sum in (\ref{spectr2}) is concerned, it gives
a vanishing contribution, since $B$ is stable  with respect to the 
strong interactions.

 We recall that by the reduction formula we obtained
the analytic continuation of the physical matrix element with respect
to the variable $k_1^2$ (the mass squared of an off-shell meson $P_1$).
Therefore, the decay amplitude can be calculated  from its discontinuity
by means of a dispersion relation in this external mass variable.
As discussed above, the integral extends along a finite interval,
 so the dispersion relation reads
\begin{equation}\label{direl}
A(m_B^2, m_1^2, m_2^2)=
{1\over \pi} \int_0^{(m_B-m_2)^2}{\mathrm d}z
 {\hbox{Im}  A(m_B^2, z+i\epsilon, m_2^2)\over z-m_1^2-i\epsilon}\,,
\end{equation}
with the discontinuity given by the first sum in Eq.~(\ref{spectr2}).

Let us see now what is the form of the dispersion relation  obtained in the 
LSZ formalism, when the analytic continuation is done with 
respect to the mass squared of the $B$ meson. To this end,  
we start again from the matrix element
$\langle P_1(k_1) P_2(k_2), out|{\mathcal H}_w(0)|B(p)\rangle$, and apply the
LSZ formula, reducing this time the initial meson $B$. We obtain
\begin{equation}\label{lsz2}
A (p^2, m_1^2, m_2^2)= {i\over \sqrt{2 \omega_B}}\int {\mathrm d}x  {\mathrm
e}^{-ipx} \theta (-x_0) \langle P_1(k_1)P_2(k_2), out|[ {\mathcal H}_w(0),
\eta _B(x)]|0\rangle\,.
\end{equation}
By exploiting the causality properties of the retarded commutator
one can prove \cite{KaWi} that the amplitude can be extended as a real
analytic function in the complex plane  $s=p^2$, with the
discontinuity across the real axis given by
\begin{eqnarray}\label{spectrs}
\hbox{Im}  A(p^2+i\epsilon, m_1^2, m_2^2)&=& {1\over 2 \sqrt{2 \omega_B}}
\sum_{n}\left[\delta  (p-p_n) \langle P_1(k_1)P_2(k_2),out|
{\mathcal H}_w|n\rangle\langle n| \eta_B|0\rangle \right. \nonumber \\
&&-\left.\delta (p_n)\langle P_1(k_1) P_2(k_2),out|
\eta_B|n\rangle\langle n|{\mathcal H}_w|0\rangle\right]\,.
\end{eqnarray}
We obtained this result in the standard way \cite{Oehme},
replacing $i\theta(-x_0)$  in (\ref{lsz2}) by $1/2$, and inserting a complete
set of states in the commutator. Actually, the second sum in
(\ref{spectrs})  vanishes because the only intermediate states
allowed have zero 4-momentum.
In the first term, the allowed particles are those connected 
to the final state through a weak process
and to $B$ through a strong transition, (we recall that the last matrix 
element vanishes when $B$ is on the mass shell). The lowest
two particle state entering the unitarity sum is $B^*\pi$ which defines 
the normal unitarity threshold.

The spectral function (\ref{spectrs}) enters a dispersion relation  of the 
form (\ref{drel}) with respect to $s=p^2$. 
However, it is obvious that such a dispersion relation is not useful for
estimating the rescattering effects in nonleptonic
$B$ decays. As discussed above, in this way one does not describe the strong
interactions in the final state, but rather the strong interactions
in the initial state.

The fact that a dispersion relation in the mass squared of the $B$ 
meson cannot describe final state rescattering effects is understood by 
simple qualitative arguments: in order to make the
analytic continuation in the variable $s$ we must reduce the $B$ meson.
Hence, the source $\eta_B$ and the weak hamiltonian ${\cal H}_w$
enter different matrix elements in the unitarity sum,
and terms describing the weak decay of
$B$ multiplied by strong scattering amplitudes cannot appear.
Therefore the procedure applied in \cite{Blok}-\cite{Falk},
based on the analytic continuation in $s$ combined with a discontinuity
containing rescattering terms is not consistent. As we pointed out above
the unitarity sum (\ref{unit}) defines the spectral function
in a dispersion relation with respect to the mass variable of one of the
final mesons. The relations (\ref{spectr2}) and (\ref{direl}) are the main 
results we obtained in the frame of the standard LSZ formalism.

A few comments about the above formulae are of interest. First, it is clear
that one can repeat the procedure by reducing the meson $P_2$ instead of $P_1$.
The corresponding expressions can be obtained easily from those given above by
permutting the indices 1 and 2. The expressions seem different, but of course
the results should be the same when a complete set of states is inserted 
in the unitarity sum.

A more subtle question, which is also connected to the completeness of the set
inserted in the unitarity sum is whether the discontinuity  defined in
(\ref{spectr2}) is real or complex. For $T$ (or $CP$) conserving interactions,
the reality of the spectral function was proved a long time ago \cite{GoTr},
 \cite{GoWa}.
It turns out that the
absorbtive part remains real even if the relevant terms in the
weak hamiltonian are not $CP$ 
conserving. We take into account the fact that these terms have
the form   
\begin{equation}\label{hweak}
{\mathcal H}_w=\sum _{j}c_j {\cal O}_j\,,
\end{equation}
where $c_j$  are complex numbers  and  ${\cal O}_j$ are products of $V$ and 
$A$ currents. Consider the spectral function
\begin{equation}\label{sigma}
\sigma (z)= \hbox{Im}  A(m_B^2, z+i\epsilon, m_2^2)\,,
\end{equation}
defined by the first sum in (\ref{spectr2}), and assume that a complete set of
$in$ states is inserted in the unitarity sum.
 Following \cite{GoTr}, \cite{GoWa} (see also \cite{Bart}) we
can express the two matrix elements in this sum as 
\begin{equation}\label{etapt}
\langle P_2(k_2) |\eta_1| n, in\rangle =
\langle P_2(k_2)|(PT)^{-1}(PT) \eta_1 (PT)^{-1}(PT)|n, in\rangle =
\langle P_2(k_2)|\eta_1|n, out\rangle ^*\,,
\end{equation}
and
\begin{equation}\label{hweakpt}
 \langle n, in | 
{\mathcal H}_w |  B(p)\rangle=
\langle n, in |(PT)^{-1}(PT) {\mathcal H}_w (PT)^{-1} (PT)| B(p)\rangle
\label{hpt1}\, =\! \langle n, out|{\mathcal H}_w| B(p)\rangle^*\,.
\end{equation}
We used here the transformation properties of the $V$ and $A$
currents under $P$ and $T$ transformations and the fact 
that under space-time reversal the particles conserve their momenta, and
the $in (out)$ states become $out (in)$ states, respectively.  
Moreover, the intrinsic parities of the states and the
operators have a product equal to $+1$, and
the matrix element are replaced by their complex conjugates,
given the antiunitary character of the operator $T$.
By using the relations (\ref{etapt}) and (\ref{hweakpt}) in
 (\ref{spectr2}) we obtain
\begin{eqnarray}\label{reality}
\sigma (z) &=& {1\over 2\sqrt{2\omega_1}}\sum_{n}\delta  (k_1+k_2-p_n)  \langle  P_2(k_2) |
\eta_1|n, in\rangle\langle n, in| {\mathcal H}_w|  B(p)\rangle \,\nonumber\\
&=&{1\over 2 \sqrt{2\omega_1}}\left [\sum_{n}\delta  (k_1+k_2-p_n)  \langle  P_2(k_2) |
\eta_1|n, out\rangle\langle n, out| {\mathcal H}_w|B(p)\rangle \right ]^*=
\sigma^*(z)\,
\end{eqnarray}
where the equivalence between the
complete sets of $in$ and $out$ states in the definition of $\sigma (z)$ 
was taken into account. From (\ref{reality}) it follows
that the discontinuity is manifestly real only if the  intermediate states 
form  a complete set. If the unitarity
sum is truncated, this property is lost, since various terms have
complex phases which do not compensate each other in an obvious way.
As noticed in \cite{GoTr}, in order to maintain the proper reality condition at
all stages of approximation, it is convenient to write the sum over the
complete set of states  $|n\rangle$ as a combination
$1/2|n, in\rangle +1/2|n, out\rangle$. This prescription
will be applied in  Section 4 when discussing the $B\to \pi\pi$ decay.

\section{Two-particle unitarity and Regge amplitudes}

In this section we write down the dispersion relation (\ref{direl}) in
the approximation that only two particle states are kept in the unitarity
sum (\ref{spectr2}).
Denoting  by $\{P_3P_4\}$ the two meson intermediate states in this sum,
the off-shell imaginary part of the decay amplitude $A_{B\to P_1P_2}$  
required in the dispersion relation (\ref{direl}) reads
\begin{eqnarray}\label{disc}
\hbox{Im}  A_{B\to P_1P_2}(m_B^2, z+i\epsilon, m_2^2)
&=&{1\over 2}\sum _{\{P_3P_4\}}\int
{{\mathrm d}^3{\mathbf k}_3\over(2\pi)^{3} 2\omega_3}{{\mathrm d}^3{\mathbf
k}_4\over(2\pi)^{3}2\omega_4}
(2\pi)^4\delta^{(4)}(p-k_3-k_4)\times\nonumber\\
&&A_{B\to P_3 P_4}(m_B^2, m_3^2, m_4^2)\, {\mathcal M}^*_{P_3 P_4 \to P_1
P_2}(s,t)\,.
\end{eqnarray}
In the sum we include the two particle states ${P_3 P_4}= { P_1P_2}$
defining the elastic channel, as well as ${P_3 P_4}\ne { P_1P_2}$ responsible
for the inelastic scattering. Let us note that the {\it c.m.}
energy is set up by the mass of the $B$ meson ($\sqrt{s}= m_B=
5.2\, \hbox{GeV})$, and  the weak decay amplitudes $ A_{B\to P_3 P_4}$
are on shell, and independent on the Mandelstam variable $t$
(or the rescattering angle $\theta$). Therefore Eq.(\ref{disc}) can be 
written as
\begin{equation}\label{sum}
\hbox {Im} A_{B\to P_1P_2}(m_B^2,z+i\epsilon, m_2^2)
=\sum _{\{P_3P_4\}} C^*_{P_3P_4;P_1P_2}(z)
A_{B\to P_3 P_4}(m_B^2, m_3^2, m_4^2)\,,
\end{equation}
where 
\begin{equation}\label{coef}
 C_{P_3P_4;P_1P_2}(z)
={1\over 2}\int
{{\mathrm d}^3{\mathbf k}_3\over(2\pi)^{3} 2\omega_3}{{\mathrm d}^3{\mathbf
k}_4\over(2\pi)^{3}2\omega_4}
(2\pi)^4\delta^{(4)}(p-k_3-k_4) {\mathcal M}_{P_3 P_4 \to P_1 P_2}(s,t)\,.
\end{equation}
These coefficients depend on the masses of all the particles participating
in the rescattering process. 
To simplify the notation we indicate explicitly only the dependence on 
the off shell mass squared  $z$ of the particle $P_1$. 
Following \cite{Dono1}-\cite{Falk} we adopt for
the strong amplitudes ${\mathcal M}$ the parametrizations obtained from 
the Regge theory \cite{Coll}
\begin{eqnarray}\label{regge}
{\mathcal M}_{P_3 P_4;P_1 P_2}(s,t)=-\sum_{V=P, f, A_2, K_2^*...} 
\gamma^V_{P_3 P_4;P_1 P_2} (t){{\mathrm
e}^{-i{\pi\alpha_V(t)\over 2}}\over \sin{\pi\alpha_V (t)\over 2}}
\left({s\over s_0}\right)^{\alpha_V(t)}\,+\nonumber\\
\sum_{V=\rho, K^*...}i
 \gamma ^V_{P_3 P_4\to P_1 P_2}(t){{\mathrm
e}^{-i{\pi
\alpha_V(t)\over 2}}\over \cos{\pi\alpha_V(t)\over 2}}\left({s\over s_0}
\right)^{\alpha_V(t)}\,,
\end{eqnarray}
where the first sum includes $C=1$ trajectories and the second one
 $C=-1$ trajectories.
As usual we take $s_0\approx 1\,\hbox{GeV}^2$ and linear trajectories
\begin{equation}\label{alpha}
\alpha_V(t)=\alpha_0+\alpha' t\,,
\end{equation}
with the standard choices \cite{Coll}
\begin{equation}\label{alphapom}
\alpha_0=1.08 \,,\quad \alpha'=0.25~{\rm GeV}^{-2} \,
\end{equation}
for the Pomeron, and
\begin{equation}\label{alphapart}
\alpha_0=0.45\,,\quad \alpha'=0.94\,\hbox{GeV}^{-2}\, \end{equation}
for all the other trajectories.
The possible divergences occuring in the expression (\ref{regge})
for $t\ne0$ are avoided by taking $\alpha(t)\approx\alpha_0$ in the
denominators. We shall therefore obtain
\begin{eqnarray}\label{aprox}
{\mathrm sin}{\pi\alpha_P(t)\over2}&\approx&1\nonumber\\
{\mathrm sin}{\pi\alpha_V(t)\over2}&\approx&{1\over\sqrt{2}},~~~
V=f,~A_2~K^*_2\nonumber\\
{\mathrm cos}{\pi\alpha_V(t)\over2}&\approx&{1\over\sqrt{2}},~~~
V=\rho,~K^*.
\end{eqnarray}
As far as the Regge  residua
$\gamma^V_{P_3 P_4;P_1 P_2} (t)$ are concerned, 
they are supposed to satisfy the
factorization relation 
\begin{equation}\label{resid}
\gamma^V_{P_3 P_4;P_1 P_2} (t)=\gamma_{P_3 P_1 V}
(t)\gamma_{P_4 P_2 V} (t)\, ,
\end{equation}
and their values 
at $t=0$ will be determined
using the optical theorem and the phenomenological Regge-like 
parametrizations of the total cross sections \cite{DoLa},\cite{Part}
(details will be given in the next Section). The $t$-dependence of these
functions is, however, poorly known and we assume they are simply constants. 

In order to perform the integral (\ref{coef}), the Mandelstam
variable $t$ is expressed in terms of the scattering angle $\theta$
\begin{equation}\label{t}
t(z)=t_0(z)+2k_{12}(z)k_{34}\cos\theta\,,
\end{equation}
with
\begin{eqnarray}\label{k12k34}
t_0(z)&=&z+m_3^2-{(m_B^2+m_3^2-m_4^2)(m_B^2+z-m_2^2)\over2m_B^2}
\nonumber\\
k_{12}(z)&=&{1\over2m_B}\sqrt{(m_B^2-z-m_2^2)^2-
4z m_2^2}\nonumber\\
k_{34}&=&{1\over2m_B}\sqrt{(m_B^2-m_3^2-m_4^2)^2-
4m_3^2m_4^2}\,,
\end{eqnarray}
where we indicated explicitely only the dependence on the variable $z$.
With these kinematic variables the integration over the momenta 
${\mathbf k}_3$ and ${\mathbf k}_4$ in (\ref{coef}) is straightforward,
and the coefficients $C_{ P_3P_4; P_1P_2}$ can be expressed as
\begin{equation}\label{coef1}
C_{ P_3P_4; P_1P_2}(z)= \sum_{V}
\gamma^V_{P_3 P_4;P_1 P_2} (0) \kappa^V_{P_3 P_4;P_1 P_2}\,,
\end{equation}
where
\begin{equation}\label{kappa}
  \kappa^V_{P_3 P_4;P_1 P_2}(z)=\xi_V { k_{34} \over 16 \pi m_B}
{\mathcal R}_V^{-1}(z)
\left[{\mathrm e}^{{\mathcal R}_V(z)}-{\mathrm e}^{-{\mathcal R}_V(z)}\right]
\exp\left[(\alpha_{0,V}+\alpha'_Vt_0)
\left(\ln {m_B^2\over s_0}-i{\pi\over2}\right)\right],
\end{equation}
\begin{equation}\label{K}
{\mathcal R}_V(z)=
2\alpha'_Vk_{12}(z)
k_{34}\left(\ln{m_B^2\over s_0}-i{\pi\over2}\right)\, ,
\end{equation}
and $\xi_V$ is a numerical factor equal to $-1$ for 
the Pomeron, $i\sqrt{2}$ for $C=-1$ trajectories and   $-\sqrt{2}$ for $C=1$  
physical trajectories.

By inserting the expression (\ref{sum}) of the spectral function in the
dispersion relation (\ref{direl}) and recalling  that the decay amplitudes 
$A_{B\to P_3P_4}$ do not depend on $z$, we obtain (for simplicity we  omit 
now the mass arguments when the amplitudes are on-shell)
\begin{equation}\label{sumint}
A_{B\to P_1P_2}=\sum_{\{P_3 P_4\}}\overline\Gamma_{P_3 P_4; P_1 P_2}A_{B\to
P_3 P_4}\,,
\end{equation}
where
\begin{equation}\label{bareta}
\overline\Gamma_{P_3 P_4;P_1 P_2}=\sum_{V}
\gamma^V_{P_3 P_4; P_1 P_2} (0) \bar\eta^V_{P_3 P_4;P_1 P_2}\,
\end{equation} 
and
\begin{equation}\label{coefint}
\bar\eta^V_{ P_3P_4;P_1P_2}={1\over \pi}
\int_0^{(m_B-m_2)^2}{\mathrm d}z{\kappa^{V*}_{ P_3P_4;
P_1P_2}(z)\over z-m_1^2-i 
\epsilon}\,.
\end{equation}
For further use we also define
\begin{equation}\label{eta}
\Gamma_{P_3 P_4; P_1 P_2}=\sum_{V}
\gamma^V_{P_3 P_4;P_1 P_2} (0) \eta^V_{P_3 P_4;P_1 P_2}\,
\end{equation} 
and
\begin{equation}\label{coefbar}
\eta^V_{ P_3P_4;P_1P_2}={1\over \pi}
\int_0^{(m_B-m_2)^2}{\mathrm d}z{\kappa^V_{ P_3P_4;
P_1P_2}(z)\over z-m_1^2-i 
\epsilon}\,.
\end{equation}
Before considering applications, let us make a few comments
about the approximations adopted when deriving  the above formulae. 
First, we notice that the Regge expression (\ref{regge}) is valid
for large $s$ and $t$ close to 0. 
The value $s=m_B^2$ satisfies this condition, but the values
of $t$ appearing in the integral upon the scattering angle 
in (\ref{coef}) can be  large, outside the range of validity of 
the Regge theory.
However, the hadronic amplitudes decrease at large $|t|$, so the contribution
of the large scattering angles in the unitarity integral is expected to be
small and not very sensitive to the inaccuracy of the dynamical model.
Another difficulty is related to the fact
that the particle $P_1$  is off shell, and its mass $k_1^2=z$ becomes very
large at the upper limit of integration in the dispersion relation
(\ref{direl}). Here again, one of the assumptions for the validity of the Regge
expression, namely $\sqrt{s}>> m_i$ \cite{Coll} is not met. However, this part
of the integral, which is not correctly evaluated, brings a small
contribution in the dispersion integral due to the denominator in
(\ref{direl}) (this statement is true when the masses of the intermediate
particles $P_3$ and $P_4$ are not too large). We emphasize that
the main advantage 
of the formalism is that it provides, with no approximation, an
algebraic relation involving only physical decay amplitudes. Indeed, as we
mentioned, all the quantities $A_{B\to P_3 P_4}$ appearing in (\ref{sumint})
are on shell, the dynamical approximations affecting only the coefficients
$\Gamma_{P_3P_4;P_1P_2}$.

\section{Constraints on the amplitudes of $B^0\to\pi^+\pi^-$ decays}

Unitarity and the dispersion relations were used in previous works 
\cite{Dono1}-\cite{Falk} in
order to  estimate the FSI corrections to the decay amplitudes calculated  in
an approximation which does not include strong rescattering (like, for
instance, factorization). Such an evaluation was made  in \cite{Falk} for the
magnitude of final state interactions in $B^-\to \pi^-\bar K^0$.
In the notations used in Section 3, it corresponds 
to $P_1P_2= \pi^-\bar K^0$, with the  intermediate states $P_3P_4=\pi^0  K^-$
and $\eta K^0$ inserted in the unitarity sum. Then only  physical
trajectories ($V=\rho,~K^*$) contribute to (\ref{regge}), for which
the  intercept and the slope are given in (\ref{alphapart}).
 
With values of the residua extracted from the phenomenological
parametrization of the cross sections the authors of Ref.
\cite{Falk} suggested a
modification of the magnitude of the decay amplitude of the decay $B^-\to
\pi^-\bar K^0$ by a factor of $10\%$. However, the
coefficients $\Gamma_{P_3P_4;P_1P_2}$ appearing in a relation of the type
(\ref{sumint})   were calculated  in \cite{Falk} by using the rescattering
absorbtive function in a dispersion relation with respect to the variable $s$
(see Ref. \cite{Falk}, Eq. 2.17).
By performing the correct calculations with the same
values of the parameters, we find that the coefficient  $\Gamma_{\pi^0  K^-;
\pi^-\bar K^0}$, for instance, is larger by a factor of about
2.5 compared 
to the value reported in \cite{Falk}.  This shows that the correct treatment
can modify the conclusions about the magnitude of FSI in $B\to\pi K$ decays.

In the present paper we consider an other application of the
dispersive formalism to the decay $B^0\to \pi^+\pi^-$. The time
dependent $CP$ asymmetry in this decay is considered as one of the ways of
extracting the angle $\alpha$ of the unitarity triangle \cite{GrLo}.
However, the unknown strong phase
difference between the tree and the penguin amplitudes of the process
affects the accuracy of this determination. Additional  theoretical 
constraints on these amplitudes would be very helpful for reducing the
uncertainty of the method. As we shall show below, the dispersion relations can
provide such a constraint. We investigate the problem by combining the
relations (\ref{sum}) and (\ref{sumint}) derived above with isospin or
$SU(3)$ symmetry \cite{Gron}-\cite{Char}. The idea is that by unitarity and
dispersion relations we obtain a set of correlations between exact decay
amplitudes, containing both weak and strong phases. By imposing in addition
$SU(3)$ flavour symmetry, all the amplitudes can be expressed in terms of a
small number of parameters, for which unitarity and the dispersion relations
provide nontrivial constraints.  Following \cite{Gron} we write most
generally the amplitude of the decay $B^0\to \pi^+\pi^-$ as a sum of diagram
contributions  
\begin{equation}\label{notation}
 A_{B^0\to \pi^+\pi^-}=
-(A_T~{\mathrm e}^{i\gamma}+A_P~{\mathrm e}^{-i\beta}+A_P'~{\mathrm
e}^{i\gamma}+A_E~{\mathrm e}^{i\gamma}+ A_{PA}~{\mathrm e}^{-i\beta})\,, 
\end{equation}
where $A_T$, $A_P (A_P')$, $A_E$  and $A_{PA}$ denote the amplitudes of the
tree, penguin, exchange and penguin annihilation 
diagrams, respectively.  We indicated explicitely the weak phases  
 defined as $\beta= {\mathrm Arg}(-V^*_{td}$) and
$\gamma= {\mathrm Arg}(-V^*_{ub})$ \cite{Part}.  As intermediate states
${P_3P_4}$ in the equations (\ref{disc}) and (\ref{sumint}) we
keep $\pi^+\pi^-$ giving the elastic channel, as well as two meson states
responsible for the soft inelastic scattering, e.g.,
$\pi^0\pi^0$, $ K^+ K^-$, $ K^0\bar K^0$, $\pi^0\eta_8$ and
$\eta_8 \eta_8$ ($\eta_8$ is the $\eta,~\eta'$ superposition
belonging to the $SU(3)$ octet). We notice that $\pi^0\eta_8$ will
not contribute finally due to isospin conservation in the strong
rescattering. 
Of course, besides these states, other inelastic channels, like
for instance multipion states can contribute.
The states $D^+ D^-$ and $D^0\bar D^0$ are not included because 
they contribute to the hard scattering
\cite{Dono1}.

Assuming SU(3) flavour symmetry we express the decay amplitudes 
$B\to P_3P_4$ of interest as \cite{Gron}-\cite{Char}
\begin{eqnarray}\label{notations}
&&A_{B^0\to \pi^0\pi^0}={1\over\sqrt{2}} (-A_C~{\mathrm
e}^{i\gamma}+ A_P~{\mathrm e}^{-i\beta}+A_P'~{\mathrm
e}^{i\gamma}+A_E~{\mathrm
e}^{i\gamma})\,,\nonumber\\
&&A_{B^0\to K^+ K^-}= -A_E~{\mathrm e}^{i\gamma}\,,\nonumber\\
&&A_{B^0\to K^0\bar K^0}=A_P~{\mathrm
e}^{-i\beta}+A_P'~{\mathrm e}^{i\gamma}\,,\nonumber\\
&&A_{B^0\to\pi^0\eta_{8}}=-{1\over\sqrt{3}}(A_P~{\mathrm
e}^{-i\beta}+A_P'~{\mathrm e}^{i\gamma}-A_E~{\mathrm e}^{i\gamma})
\,,\nonumber\\ 
&&A_{B^0\to\eta_{8}\eta_{8}}={1\over3\sqrt{2}}(A_C~{\mathrm
e}^{i\gamma} +A_P~{\mathrm e}^{-i\beta}+A_P'~{\mathrm
e}^{i\gamma}+A_E~{\mathrm e}^{i\gamma} )\,,
\end{eqnarray}
where $A_T$, $A_P$, $A_P'$ and $A_E$  are the same as in (\ref{notation})
and $A_C$ denotes the amplitude of the tree colour suppressed  diagrams. 

Due to the lack of detailed dynamical calculations, 
various phenomenological assumptions are made in the literature 
about the above amplitudes. The conservative bound 
$\vert A_P/A_T \vert <1$ for the ratio of the penguin and tree amplitudes is
mentioned in  \cite{Char} 
(a more specific estimate $\vert A_P/A_T \vert \approx 0.2$ is also quoted 
in this reference). The penguin annihilation amplitude $A_{PA}$ correspond to 
OZI-suppressed diagrams \cite{Char}, while $A_C$ and $A_E$ are colour suppresed by
a factor of about $0.25$ with respect to the corresponding colour favoured
amplitude. Finally, the two penguin amplitudes $A_P$ and $A_P'$  are assumed
to satisfy  $\vert A_P'/A_P \vert \approx 0.4$ \cite{Char}. Using these
estimates, we assume as a first approximation that we can neglect in
(\ref{notation}) and (\ref{notations}) the suppressed terms  $A_P'$, $A_C$,
$A_E$ and $A_{PA}$, keeping only the dominant amplitudes $A_T$ and $A_P$. 

The next step is to introduce the decay amplitudes $A_{B\to P_iP_j}$ discussed
above  in the dispersion relation (\ref{sumint}). 
We should recall however that, due to the truncation of the unitarity sum, 
the imaginary part of the decay amplitude $A_{B\to P_1P_2}$ 
obtained  from  (\ref{sumint}) (or equivalently  from the unitarity relation
(\ref{sum}) evaluated on-shell) might be not real. In order to avoid this
situation we apply the procedure suggested in Ref.\cite{GoTr} which
maintains the proper reality condition of the spectral function and simulates
the effect of other inelastic channels. As discussed at the end of Section
2, this method amounts to insert in the unitarity sum the complete set of
states $1/2|in\rangle+1/2|out\rangle$.  We recall that this method was
applied  to include inelastic effects  through complex phases in the
dispersive analysis of the electromagnetic form factors \cite{Omnes},
\cite{Bart}. In our case this procedure yields, instead of (\ref{sumint}),
the modified  dispersion relation
\begin{equation}\label{sumint1}
 A_{B\to P_1P_2}
={1\over 2}\sum _{\{P_3P_4\}} \Gamma_{P_3P_4;P_1P_2}
A^*_{B\to P_3 P_4}+ {1\over 2}\sum_{\{P_3P_4\}} \bar\Gamma_{P_3P_4;P_1P_2}
A_{B\to P_3 P_4}\,,
\end{equation}
where $\Gamma_{P_3P_4;P_1P_2}$ and $\overline\Gamma_{P_3P_4;P_1P_2}$ 
are defined in  (\ref{eta}) and (\ref{bareta}), respectively. 
We notice that Eq.(\ref{sumint1}) can be splitted in 
two relations, one for the real part and another for the 
imaginary part of the decay amplitude. In particular, the relation giving
the imaginary part is  \begin{equation}\label{sum1} i A^*_{B\to 
P_1P_2}- iA_{B\to P_1P_2} =\sum _{\{P_3P_4\}}
C_{P_3P_4;P_1P_2}(m_\pi^2)A^*_{B\to  P_3 P_4}+  \sum_{\{P_3P_4\}}
C^*_{P_3P_4;P_1P_2}(m_\pi^2) A_{B\to P_3 P_4}\,,  \end{equation} 
and can be obtained  also directly 
from the unitarity relation  (\ref{sum}) evaluated on shell. Concerning 
the real part, it is obtained by taking the principal
value of the dispersion integrals appearing in (\ref{coefint}) and
(\ref{coefbar}).

We describe now briefly the determination of 
the Regge residua $\gamma^V_{P_3 P_4;P_1 P_2}(0)$ which enters 
the expressions (\ref{bareta})  and (\ref{eta}) of
the coefficients of the dispersion relation (\ref{sumint1}).
 We use the optical theorem and the Regge parametrization of the total
hadronic cross-sections \cite{DoLa},\cite{Part}, which gives
\begin{equation}\label{optical}
{s_0\over s}~{\rm Im}{\mathcal M}_{f\to f}(s,0)\approx
s_0\sigma_{tot} 
=s_0X\left({s\over s_0}\right)^{\alpha_P(0)-1}+s_0Y
\left({s\over s_0}\right)^{\alpha(0)-1}
\end{equation}
where $s_0\approx 1 {\rm GeV}^2\approx{1\over0.38}{\rm mb}^{-1}$.
The first term  represents the Pomeron
contribution, the second the contributions of all the other trajectories. 
By comparing (\ref{optical}) with the Regge parametrization (\ref{regge}) we 
obtain $$s_0X=\gamma^P_{f;f}\,,\,\,\,\, s_0Y=\sum_{V\ne P}\gamma^V_{f;f}\,.$$

For the Pomeron, which contributes to the elastic $\pi^+\pi^-$ channel, we
assumed  that the coupling constant is proportional to the number of
quarks, taking as in \cite{Dono1} 
$\gamma^P_{\pi^+\pi^-;\pi^+\pi^-}=\left({2\over3}\right)^2s_0X_{NN}$.   
The residua of the physical trajectories (which in our case are: $\rho,
f,f'K^*, K_2^*$ and $A_2$) were estimated by taking into account the
factorization property  (\ref{resid}) combined with the experimental data on
several hadron-hadron scattering processes. We used the processes given in
Table~\ref{t:table1}, for which we wrote the contributions of various
trajectories as in \cite{Coll}. 
\begin{table}
\begin{center}
\begin{tabular}{c|l} \hline &\\
Process&Contributing trajectories\\
&\\
\hline
&\\
$\pi^-p$&$P+f+f'-\rho$\\
$\pi^+p$&$P+f+f'+\rho$\\
$\bar p p$&$P+f+f'+\rho+\omega+\phi+A_2$\\
$\bar p n$&$P+f+f'-\rho+\omega+\phi-A_2$\\
$ p p$&$P+f+f'-\rho-\omega-\phi+A_2$\\
$p n$&$P+f+f'+\rho-\omega-\phi-A_2$\,
\end{tabular}
\end{center}
\caption{Contributing trajectories to various hadronic processes.}
\label{t:table1}
\end{table}
\noindent
Using the $NN$ channels we write in particular
\begin{eqnarray}
&&s_0Y_{pp}=\gamma^2_{N\bar Nf}-\gamma^2_{N\bar
N\rho}-\gamma^2_{N\bar N\omega}
-\gamma^2_{N\bar N\phi}+\gamma^2_{N\bar NA_2}\nonumber\\
&&s_0Y_{pn}=\gamma^2_{N\bar Nf}+\gamma^2_{N\bar
N\rho}-\gamma^2_{N\bar N\omega}
-\gamma^2_{N\bar N\phi}-\gamma^2_{N\bar NA_2}.
\end{eqnarray}
Noticing that experimentally $Y_{pp}\approx Y_{pn}$ \cite{Part} we obtain
\begin{equation}
\gamma^2_{N\bar NA_2}\approx \gamma^2_{N\bar N\rho}\,
\end{equation}
and further
\begin{equation}
s_0(Y_{\bar pp}-Y_{\bar pn})=2\gamma^2_{N\bar
N\rho}+2\gamma^2_{N\bar NA_2}\approx4\gamma^2_{N\bar
N\rho}\,.
\end{equation}
Also, by replacing the contributions of $f$ and $f'$ with the octet member
$f_8$, we  obtain 
\begin{eqnarray}\label{Y}
&&s_0(Y_{\pi^-p}-Y_{\pi^+p})=2\gamma_{\pi^+\pi^-\rho}~\gamma_{N\bar
N\rho}\nonumber\\
&&s_0(Y_{\pi^-p}+Y_{\pi^+p})=2\gamma_{\pi^+\pi^-f_8}~\gamma_{N\bar
Nf_8}\nonumber\\ &&s_0(Y_{pn}+Y_{pp}+Y_{\bar pn}+Y_{\bar
pp})=4\gamma^2_{N\bar Nf_8}. \end{eqnarray}
The coupling constants $\gamma^2_{\pi^+\pi^-\rho^0}$ and
$\gamma^2_{\pi^+\pi^-f_8}$ can be easily calculated from these
equations using the experimental values of the parameters 
$X$ and $Y$ for $\pi N$ and $NN$ scattering \cite{Part}. 
Other coupling constants we need are obtained from
the previous ones by using $SU(3)$ symmetry, namely
\begin{eqnarray}
&&\gamma^2_{\pi^0\pi^-\rho^-}=\gamma^2_{\pi^+\pi^-\rho^0}
\nonumber\\
&&\gamma^2_{\bar
K^0\pi^-K^{*-}}={1\over2}\gamma^2_{\pi\pi\rho}\nonumber\\
&&\gamma^2_{\bar K^0\pi^-K_2^{*-}}={3\over2}\gamma^2_{\pi^+\pi^-f_8}
\nonumber\\
&&\gamma^2_{\pi\eta_8A_2}=\gamma^2_{\pi^+\pi^-f_8}\,.
\end{eqnarray}

In Table~\ref{t:table2} we give the values of the Regge residua
calculated with the aid of the above relations using the
parameters of the total cross sections quoted in \cite{Part},
the coefficients $\kappa^V_{P_3P_4;P_1P_2}$ are defined in
(\ref{kappa}), while the coefficients
$\bar\eta^V_{P_3P_4;P_1P_2},~\eta^V_{P_3P_4;P_1P_2}$ are given by
Eqs. (\ref{coefint}), and (\ref{coefbar}) respectively. We notice that the
dominant contribution is brought by the elastic channel, and in particular by
the Pomeron  exchange.
\begin{table}
\begin{tabular}{c|c|c|c|c|c}
\hline
&&&&&\\
$P_3~P_4$&$V$&$\gamma^V_{P_3P_4}$&$\kappa^V_{P_3P_4;P_1P_2}$&
$\eta^V_{P_3P_4;P_1P_2}$&$\bar\eta^V_{P_3P_4;P_1P_2}$\\
&&&&&\\
\hline
$\pi^+\pi^-$&$P$&25.6&$-0.0089+0.0270 i$&$-0.0547+0.0768 i$&
$-0.0007-0.0946 i$\\
$\pi^+\pi^-$&$\rho$
&31.4&$-0.0005-0.0015
i$&$0.0001-0.0051 i$&$-0.0029+0.0040 i$\\
$\pi^+\pi^-$&$f_8$&35.3&$-0.0015+0.0005
i$&$-0.0051-0.0001 i$&$-0.0040-0.0029 i$\\
$\pi^0\pi^0$&$\rho$&
31.4&$-0.0005-0.0015
i$&$0.0001-0.0051 i$&$-0.0029+0.0040 i$\\
$\bar K^0~K^0$&$K^*$&15.7&$-0.0003-0.0007
i$&$-0.0004-0.0030 i$&$-0.0019+0.0024 i$\\
$\bar K^0~K^0$&$K^*_2$&52.9&$-0.0007+0.0003 i$&
$-0.0030+0.0004 i$&$0.0024-0.0019 i$\\
$\eta_8\eta_8$&$A_2$&$35.3$&$-0.0015+0.0005i$&
$-0.0051-0.0001i$&$-0.0040-0.0029i$\\
\hline
\end{tabular}
\caption{The values of the Regge residua and of the coefficients entering
the dispersion relation.}
\label{t:table2}
\end{table}

With the numbers given in Table~\ref{t:table2}, the real and 
imaginary parts of the dispersion relation (\ref{sumint1}) can be
completely evaluated, yielding two algebraic relations for the decay
amplitudes written in (\ref{notation}) and (\ref{notations}). According to the
discussion following these equations, we shall neglect in the first 
approximation the suppressed diagrams, keeping only the   contribution of 
the tree and the penguin  diagrams $A_T$ and $A_P$. 
Both these amplitudes have in principle strong  rescattering phases, 
which we denote as $\delta_T$ and $\delta_P$, respectively. Let us write
\begin{equation}\label{phase} 
 {A_P\over A_T}=R{\mathrm e}^{i\delta}\,, 
\end{equation}  
where $R = \vert A_P/ A_T \vert$ and
$\delta=\delta_P-\delta_T$. For simplicity, let us make also the additional
assumption that the tree amplitude $A_T$ is real. Then this amplitude  can be
factored out  from  the dispersion relation (\ref{sumint1}), which gives the 
following two constraints for the ratio $R$ and the phase difference $\delta$ 
of the penguin and tree amplitudes describing the $B\to \pi^+\pi^-$ decay
\begin{eqnarray}\label{final1}  
1.1189 {\rm sin}(\gamma+0.721)+
R{\rm sin}(\delta-\beta+0.686)=0\,, \nonumber \\ 
-1.239{\rm sin}(\gamma-0.0403)+R{\mathrm  sin}(\delta-\beta+1.452)= 0\, .
\end{eqnarray}  
These equations  can be explicitely solved  as
\begin{eqnarray}\label{final2} 
\delta &=& \beta +\epsilon(\gamma)\,,\nonumber\\ ~\nonumber \\ 
R &=& -1.1189{\sin{(\gamma + 0.721)}
\over \sin{(\epsilon(\gamma) + 0.686)}}\,,  \end{eqnarray} 
where
\begin{equation}\label{final3}
\epsilon(\gamma) = -\arctan \left[ {\sin{(\gamma - 0.0403)}\sin{0.686}
+0.9027\sin{(\gamma + 0.721)}\sin{1.452} \over \sin{(\gamma -0.0403)}\cos{0.686}
+0.9027\sin{(\gamma + 0.721)}\cos{1.452}} \right]\,.
\end{equation}
We recall that in these relations  
$\beta$ and $\gamma$ are the angles of the unitarity triangle 
which are expected \cite{Fleisch}-\cite{Parodi} to be  in the ranges
$0.17\leq\beta\leq0.52$ and $0.349\leq\gamma\leq2.79$.

In Fig.\ref{fi:phase} and Fig.\ref{fi:ratio} we represent the strong phase
difference  $\delta$ as a function of $\gamma$ for two values of
$\beta$ at the limits of the allowed intervals mentioned above, and the ratio
$R$ as a function of $\gamma$, according to (\ref{final2}). 
One can see that despite the crude approximations we made, the results are  
qualitatively reasonable. We  notice that the equation for  $\delta$ is not 
restrictive for the weak angles, while the expected condition $R<1$  
is satisfied only for a small range above $\gamma=2.4$. 
However, this somewhat intriguing limitation disappears if we relax the last
approximation made above, namely that the tree amplitude $A_T$ is real. It can
be easily seen that by allowing a nonzero $\delta_T$ in the dispersion
relation we get two equations of the form (\ref{final1}), with
$\gamma$  replaced by $\gamma+\delta_T$ and $\delta$ replaced by $\delta_P$.
The two constraints similar to  (\ref{final1}) involve now three parameters,
$R$, $\delta_T$ and $\delta_P$.      

\section{Conclusions}  
\begin{figure}
\epsfxsize=14cm
\centerline{\epsfbox{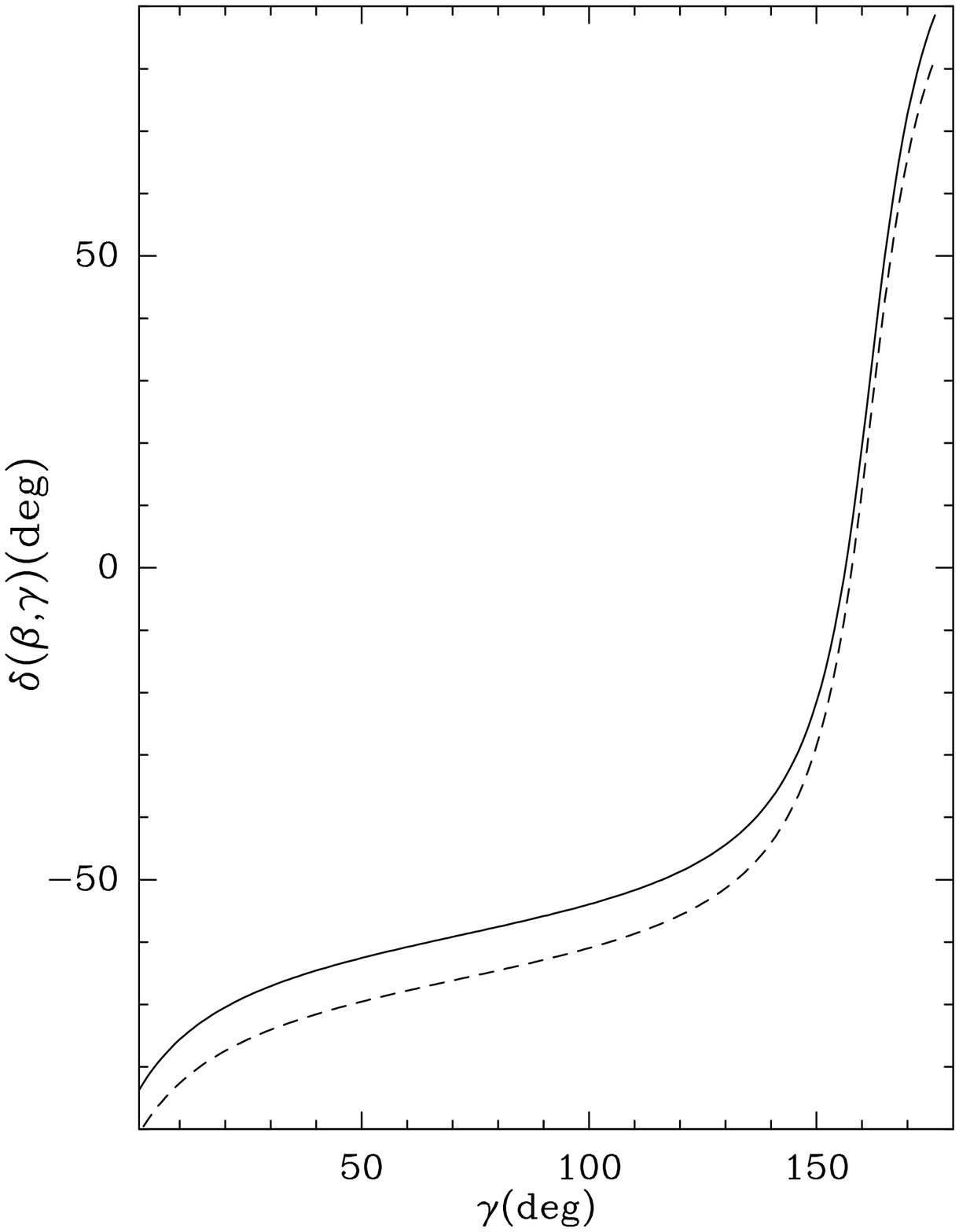}}
\caption{The strong phase difference as function of $\gamma$,
solid curve $\beta = 1\deg$, dashed curve $\beta = 174\deg$.}
\label{fi:phase}
\end{figure}

\begin{figure}
\epsfxsize=14cm
\centerline{\epsfbox{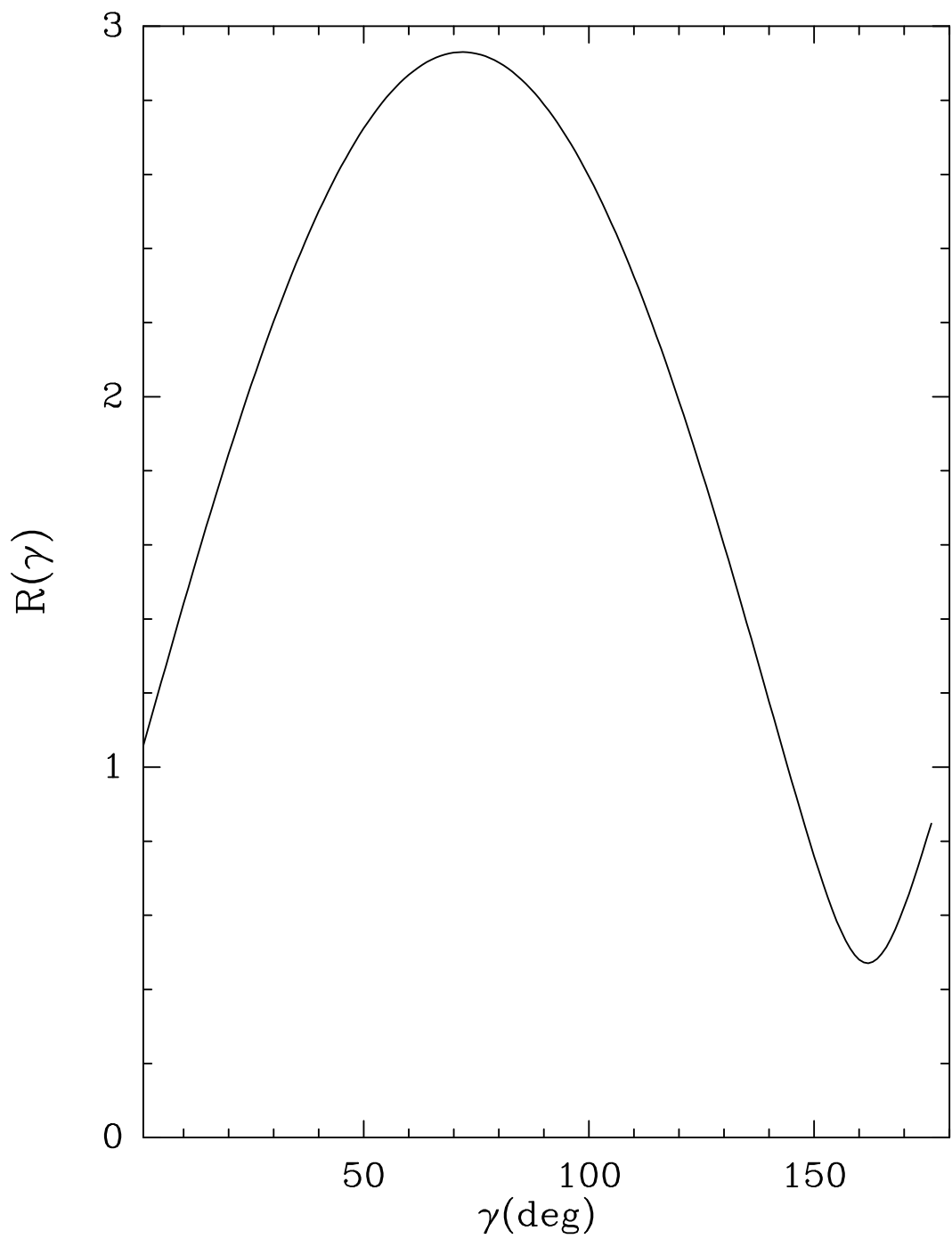}}
\caption{The ratio R as function of $\gamma$.}
\label{fi:ratio}
\end{figure}

In the present paper we investigated a recent
treatment of the final state interactions  in the nonleptonic $B$ decays,
based on unitarity and dispersion relations \cite{Dono1}-\cite{Falk}. By
considering the analytic continuation in the external mass variable in the
frame  of LSZ formalism, we established the connection between the dispersion
variable and the part of the unitarity sum defining the spectral function.
The strong rescattering part is shown to appear as a discontinuity in a
dispersion relation  in terms of the mass of one final particle. Our results
prove that the dispersion relations written in \cite{Dono1}-\cite{Falk},
based on the analytic continuation in the mass of $B$,  are not consistent.
We derived the correct dispersion relation, and 
showed that it modifies the conclusions of \cite{Falk} on the magnitude of FSI
effects in $B\to \pi K$ decay by a factor of approximately 2.5. 
We also applied the formalism to derive a theoretical
constraint for  the amplitudes of the $B^0\to\pi^+\pi^-$ decay. We included
in the unitarity sum a few channels, connecting them by the
$SU(3)$ symmetry \cite{GrLo},\cite{Gron} and took into account
qualitatively the effect of higher inelastic channels by a
procedure applied in the study of the electromagnetic form
factors \cite{GoWa}, \cite{Omnes}.
In spite of  the various dynamical assumptions mentioned in the text
our results (\ref{final2}) for the ratio $R$ and the strong relative phase
$\delta$ of the penguin and tree amplitudes in terms of the weak angles
(represented in Figs. \ref{fi:phase} and \ref{fi:ratio}) are qualitatively 
reasonable. As we mentioned, these results can be immediately modified to
incorporate a nonzero strong phase for the tree amplitude. Also, the
contributions of the suppresed diagrams neglected in the present analysis, as
well as corrections to the exact $SU(3)$ symmetry, can be easily included  in
the dispersion formalism. A more complete analysis will be made in a future
work. The results might be useful as additional constraints in the extraction
of the angles of the unitarity triangle from the time dependent $CP$
asymmetry in the decay $B^0\to\pi^+\pi^-$. 
\vskip 0.5cm  {\bf Acknowledgements:}  Two of the
authors (I. C. and L. M.) express their thanks to the Center of Particle
Physics (CPPM) and the Center of Theoretical Physics of Marseille for
hospitality.  Useful discussions with the members of the ATLAS group of CPPM
are gratefully acknowledged. This work was partly
realized in the frame of the Cooperation Agreement between IN2P3
and NIPNE-Bucharest and the Agreement between CNRS and the
Romanian Academy.

\newpage
\newpage
\end{document}